\RequirePackage{lineno}

\documentclass[structabstract]{aa}

\usepackage{color}
\usepackage{soul}
\sethlcolor{yellow}
\usepackage{txfonts}
\usepackage{natbib}
\usepackage{graphicx}
\usepackage{amsbsy}
\bibpunct{(}{)}{;}{a}{}{,} 

\begin{document}


  \title{Relativistic Outflow Drives $\gamma$-Ray Emission in 3C~345}

  \authorrunning{Schinzel et al.}
	
  \author{
    Frank~K.~Schinzel$^{(1)}$\thanks{Member of the International
Max Planck Research School (IMPRS) for Astronomy and Astrophysics at the
Universities of Bonn and Cologne.} \and
    Andrei~P.~Lobanov$^{(1)}$ \and
    Gregory~B.~Taylor$^{(2,4)}$ \and
    Svetlana~G.~Jorstad$^{(3,5)}$ \and
    Alan~P.~Marscher$^{(3)}$ \and
    J.~Anton~Zensus$^{(1)}$
  }

  \institute{
    \inst{1}~Max-Planck-Institut f\"ur Radioastronomie, Auf dem H\"ugel 69, 53121 Bonn, Germany\\
    \inst{2}~Department of Physics and Astronomy, University of New Mexico, Albuquerque NM, 87131, USA\\
    \inst{3}~Institute for Astrophysical Research, Boston University, Boston, MA 02215, USA\\
    \inst{4}~also an adjunct astronomer at the National Radio Astronomy Observatory\\
    \inst{5}~St.Petersburg State University, St.Petersburg, Russia
  }

	\date{\textit{accepted by A\&A:} 29/10/2011}

	\abstract{}
	{3C\,345 was recently identified as a $\gamma$-ray
emitter, based on the first 20 months of \textit{Fermi}-LAT data and
optical monitoring. In this paper, a connection between the
$\gamma$-ray and optical variability of 3C\,345 and properties of its
parsec-scale radio emission is investigated.}
	{The \textit{Fermi}-LAT data of 3C\,345, covering an energy range of 0.1-300\,GeV, were
combined with 32 Very Long Baseline Array observations of the object
made at 43.2\,GHz in the period of January 2008 - March 2010.}
	{The VLBA data reveal morphology and kinematics of the flow on scales of up
to $\approx 5$ milliarcseconds (deprojected linear distances of 380
parsecs).  The brightness temperature, $T_\mathrm{b}(r)$, measured
along the jet first decreases with distance $\propto r^{-(0.95\,\pm\,0.69)}$ and
later exhibits a break at $\approx 0.3$ milliarcseconds (mas), with
$T_\mathrm{b}(r) \propto r^{-(4.11\,\pm\,0.85)}$ at larger separations. Variations
of the $\gamma$-ray, optical and parsec-scale radio emission show a
similar long-term trend persistent during the entire VLBA monitoring
period. The $\gamma$-ray and optical variations on shorter time scales
are related to structural changes in the jet on scales of $\approx
0.3$\,mas ($\approx 23$\,parsecs, deprojected), with the $\gamma$-ray
and optical flares possibly related to the evolution of four distinct
superluminal components identified in the flow.} 
	{The observations indicate that both the quiescent and flaring components of the
$\gamma$-ray emission are produced in a region of the jet of $\sim
23$\,pc in extent. This region may mark the Compton-loss dominated
zone of the flow and its large extent may favor the synchrotron
self-Compton mechanism for $\gamma$-ray production in the relativistic jet
of the quasar 3C\,345.}

	\authorrunning{Schinzel et al.}

	\maketitle
	
	\section{Introduction}

	Observations of high-energy emission (keV -- TeV range) provide important 
insights into the physical properties of Active Galactic Nuclei (AGN), in particular
for the subgroup of quasars. However, high energy production scenarios are still heavily debated
with many possibilities discussed \citep[e.g.,][]{2009ApJ...703.1168B, 2009ApJ...692...32D, 2004ApJ...600L..27B}. 
Most of these physical models depend on the emission site, mainly its distance to the central engine. A connection between 
cm-/mm-radio and high energy emission of quasars was already suggested by \citet{1996A&AS..120C.491V}, 
\citet{2001ApJS..134..181J} and references therein. 
To investigate this connection, variability of AGN are studied in coordinated multi-wavelength observation
campaigns covering radio to $\gamma$-ray wavelengths. Their results, especially 
on archetypal sources, are essential for the physical understanding of the 
emission processes in AGN up to the highest energies \citep[cf. ][]{Longair2011}. 
These observations, in particular simultaneous high-resolution 
very long baseline interferometry (VLBI) observations at mm-wavelengths, provide 
new limits on the parameter space of proposed emission models, placing constraints 
on size, energetics and locations of emission regions.
 
The archetypal quasar \object{3C\,345} is one of the best studied ``superluminal'' radio sources, 
with its parsec-scale radio emission monitored over the past 30 years, in particular by
VLBI ({\em e.g.}:
\citealt{1986ApJ...308...93B}; \citealt{1992A&A...257...31B}; \citealt{1996PhDT........91L};
\citealt{1995ApJ...443...35Z}; \citealt{2000A&A...354...55R};
\citealt{2004PhDT........04K}; \citealt{2007AJ....134..799J}). Based on previous
observations, spectral index and turnover frequency distributions were obtained
\citep{1996PhDT........91L, 1998A&AS..132..261L, 2000A&A...354...55R}, the spectral
evolution of the jet was studied \citep{1999ApJ...521..509L}, properties of
radio and X-ray emission were related \citep{1997ApJ...480..596U,2005A&A...431..831L}, 
the dynamics of the central region was investigated \citep{2005A&A...431..831L} and opacity 
in the nuclear region was used to determine the physical properties and matter
composition of the compact jet \citep{1998A&A...330...79L, 2005ApJ...619...73H}.
Quasi-periodic variations of emission have been detected in the optical
\citep{1984Ap.....20..461B,1990A&A...239L...9K} and radio
\citep{1996ASPC..110..208A, 1998A&AS..132..305T, 1999ApJ...521..509L} regimes
with a possible periodicity of 3.5--4.5 years and major
flares occurring every 8--10 years. A new cycle of such enhanced nuclear
activity began in early 2008 \citep{2009ATel.2222....1L}. 

The identification
of 3C\,345 as a $\gamma$-ray source was unclear during the EGRET-era \citep[1991--2000; ][]{1993ApJS...86..629T} with a possible
weak detection between April and May, 1996 \citep{2008A&A...489..849C}.
The $\gamma$-ray emission detected by the Large Area Telescope (LAT), on board the \textit{Fermi} satellite, in the vicinity of 
3C\,345 was initially associated with another quasar in the region, B3\,1640+396 with low confidence, 
based on the 3-month data (August -- October, 2008) collected 
\citep{2009ApJ...700..597A}. However, with 20 months (August 2008 -- April 2010) of LAT monitoring data, 3C\,345 was identified as 
a $\gamma$-ray source at GeV energies, based on multi-wavelength counterpart localizations and correlated variability 
\citep{Schinzel_2010A}. 

	This paper presents results from a coordinated observational campaign targeting 3C\,345 and combining the 20-month
\textit{Fermi}-LAT $\gamma$-ray monitoring data \citep{Schinzel_2010A} with monthly VLBI
observations made at 43.2\,GHz (7\,mm wavelength) at the VLBA\footnote{Very Long 
Baseline Array of the National Radio Astronomy
Observatory, Socorro, USA}. Section 2 discusses the observations and data reduction
methods applied to obtain calibrated datasets for further analysis. It discusses in 
particular the criteria that were developed in order to determine the statistical significance 
of the 2D-Gaussian models (modelfits)
applied to the observed radio brightness distribution. In Section 3 we present 
kinematics of the pc-scale radio jet and flux density evolution, and a 
possible connection with the high energy $\gamma$-ray emission. Section 4 discusses
the findings of this paper and in Section~5 conclusions are made from the results presented here.  

	Throughout this paper a flat $\Lambda$CDM cosmology is assumed, with
$H_0 = 71$\,km\,s$^{-1}$\,Mpc$^{-1}$ and $\Omega_\mathrm{M}$ = 0.27. At the
redshift $z=0.593$ \citep{1996ApJS..104...37M} of 3C\,345 this relates to a
luminosity distance $D_\mathrm{L} = 3.47$\,Gpc, a linear scale of 6.64 pc
mas$^{-1}$ and a proper motion scale of 1\,mas\,year$^{-1}$ corresponding
to 34.5\,c.

	\section{Observations \& Data Analysis}
	
	\subsection{Fermi-LAT}

	The \textit{Fermi}-LAT \citep{2009ApJ...697.1071A} is a pair conversion
telescope designed to cover the energy band from 20 MeV to greater than 300~GeV.
It is the product of an international collaboration between NASA and DOE in the
U.S. and many scientific institutions across France, Italy, Japan, and Sweden.

	The $\gamma$-ray emission of 3C$\,$345 was identified based on correlations
found between the optical variability and major $\gamma$-ray events observed by
\textit{Fermi} LAT between August 2008 and April 2010. The $\gamma$-ray
counterpart of 3C\,345 was localized to R.A. 16$^\mathrm{h}$43$^\mathrm{m}$0.24$^\mathrm{s}$,
Dec. $+$39\degr48\arcmin22.7\arcsec \citep{Schinzel_2010A}. For this paper 
a light curve for which the $\gamma$-ray monitoring data was split into regular time intervals, each integrating
over periods of 7 days and an energy range of 0.1-300\,GeV, was obtained in the fashion described in 
\citet{Schinzel_2010A}. The position of the $\gamma$-ray counterpart was fixed to the radio localization
of 3C\,345. For the spectral shape of the $\gamma$-ray emission of 3C\,345 a power-law was used with 
the spectral index fixed to its 20 month average value of $\Gamma=2.45$\footnote{The photon spectral index $\Gamma$ is defined 
as N(E)\,$\propto$\,E$^{-\Gamma}$, where N(E) is the $\gamma$-ray photon flux as a function of energy E.}. 
The particular time binning of 7 days in this case 
provides the best trade-off between time resolution and signal to noise. This yielded a light curve with 81 significant 
detections and five 2$\sigma$ upper limit time intervals (JD\,2454756, 2454826, 2454903, 2454945, 2455155), in total this 
covers a time period of 602 days (20 months). In order to homogenize the light curve, 2$\sigma$ 
upper limits were used as values with their error estimate for that interval replaced 
with half the difference between that upper limit and its value determined through the unbinned spectral
likelihood analysis. This method was applied for the calculation of the variability index in \citet{Schinzel_2010A} and
\citet{2010ApJS..188..405A}.

	\subsection{Very Long Baseline Array (VLBA) \label{sec:vlbiobs}}

	Following the onset of a new period of flaring activity in 2008, a dedicated 
monthly monitoring campaign was initiated, using the VLBA to monitor the radio emission 
of 3C\,345 at 43.2, 23.8, and 15.4\,GHz (VLBA project codes: BS193, BS194). In this
paper only the 43.2\,GHz observations are discussed, while the analysis of 15.4 and 23.8\,GHz
data is continued. The observations were made with a bandwidth of 32\,MHz (total 
recording bit rate 256\,Mbits\,s$^{-1}$). A total of 12 VLBA observations were completed, 
with about 4.5 hours at 43.2\,GHz spent on 3C\,345 during each observation. Scans on 
\object{3C\,345} were interleaved with observations of \object{J1310+3233} (amplitude check, EVPA
calibrator), \object{J1407+2827} (D-term calibrator), and \object{3C\,279}
(amplitude check, EVPA calibrator). The VLBA data were correlated at the NRAO
VLBA hardware correlator and starting from December 2009 the new 
VLBA-DiFX correlator was employed. Analysis was done with NRAO's Astronomical Image Processing
System (AIPS) and Caltech's Difmap \citep{1995BAAS...27..903S} software for
imaging and modeling. Corrections were applied for the parallactic angle and for
Earth's orientation parameters used by the VLBA correlator. Fringe fitting was
used to calibrate the observations for group delay and phase rate. A summary
of all the observations is presented in Table~\ref{tab:obssummary}. Here, the
data from the 12 epochs of this dedicated monitoring campaign are presented,
combined with 20 VLBA observations from the blazar monitoring program of Marscher et al. (VLBA
project codes BM256, S1136) available
online\footnote{http://www.bu.edu/blazars/VLBAproject.html}. The combined
data (see Table~\ref{tab:obssummary}) cover a period
from January 2008 to March 2010, with observations spaced roughly at
monthly intervals or shorter.

	\begin{table}
 		\caption{Summary of VLBA Observations.}
		\label{tab:obssummary}
		\begin{center}
		\begin{tabular}{c c c c c}
 			\hline\hline
			Date & $S_\mathrm{tot}$ & $D$ & Beam (bpa) &  Ref. \\
			     & [Jy] &   & [mas]$\times$[mas] (\degr)   & \\
			\hline
			2008-01-17 & 1.90 & 2300 & 0.31$\times$0.19 (-27.2) &
1\\
			2008-02-29 & 2.02 & 1600 & 0.37$\times$0.21 (-29.1) &
1\\
			2008-06-12 & 2.44 & 2100 & 0.38$\times$0.16 (-28.2)
 & 1\\
			2008-07-06 & 2.23 & 800  & 0.33$\times$0.15 (-16.6) &
1\\
			2008-08-16 & 3.78 & 2700 & 0.41$\times$0.19 (-30.2) &
1\\
			2008-09-10 & 3.67 & 4300 & 0.37$\times$0.19 (-32.5) &
1\\
			2008-11-16 & 4.43 & 300 & 0.37$\times$0.33 (-1.49) &
1\\
			2008-12-21$^\dag$ & 2.91  & 5500 & 0.39$\times$0.17
(-16.9) & 1\\
			2009-01-24 & 5.04 & 8900 & 0.31$\times$0.17 (-20.5) &
1\\
			2009-02-19$^\dag$ & 3.55 & 2200 & 0.37$\times$0.20
(-19.0) & 2\\
			2009-02-22 & 4.63 & 8800 & 0.35$\times$0.15 (-19.1) &
1\\
			2009-03-16 & 6.01 & 2400 & 0.43$\times$0.30 (9.31) & 2\\
			2009-04-01 & 5.67 & 8000 & 0.33$\times$0.16 (-18.7) &
1\\
			2009-04-21 & 5.78  & 2600 & 0.33$\times$0.22 (-16.7) &
2\\
			2009-05-27 & 7.00 & 3500 & 0.33$\times$0.18 (-15.2) &
2\\
			2009-05-30 & 8.65 & 6900 & 0.32$\times$0.16 (-19.8)
& 1\\
			2009-06-21 & 5.04 & 6500 & 0.28$\times$0.16 (-10.9) &
1\\
			2009-06-29 & 6.43 & 2600 & 0.29$\times$0.16 (-11.6) &
2\\
			2009-07-27 & 7.63 & 2900 & 0.38$\times$0.16 (-30.8) &
2\\
			2009-08-16 & 5.80 & 4300 & 0.30$\times$0.16 (-20.4) & 1\\
			2009-08-26 & 6.58 & 2000 & 0.32$\times$0.17 (-24.8) &
2\\
			2009-09-16 & 6.75 & 4600 & 0.35$\times$0.19 (18.3) & 1\\
			2009-10-01 & 6.67 & 2300 & 0.22$\times$0.18 (-21.6) &
2\\
			2009-10-16 & 6.80 & 4800 & 0.41$\times$0.17 (-30.5) & 1\\
			2009-11-07 & 6.38 & 1400 & 0.32$\times$0.16 (-15.3) &
2\\
			2009-11-28 & 5.31 & 4900 & 0.28$\times$0.16 (-24.6) & 1\\
			2009-11-30 & 5.53 & 2000 & 0.29$\times$0.16 (-16.9) &
2\\
			2009-12-28 & 4.95 & 2400 & 0.31$\times$0.16 (-8.14) &
2\\
			2010-01-10 & 4.25 & 2400 & 0.33$\times$0.20 (-8.75) & 1\\
			2010-02-11 & 5.57 & 2100 & 0.27$\times$0.15 (-17.0) & 1\\
			2010-02-15 & 5.11 & 1300 & 0.31$\times$0.20 (-6.37) & 2\\
			2010-03-06 & 5.38 & 3300 & 0.39$\times$0.28 (-12.1) & 1\\
			\hline
		\end{tabular}\\[12pt]
		\end{center}
		\begin{flushleft}
		{\bf Notes:} $S_\mathrm{tot}$ -- total flux density recovered in
VLBA image;\\
		$D$ -- dynamic range measured as a ratio of the image peak flux
density to the RMS noise;\\
		Beam (bpa) -- beam size, major axis vs minor axis with position
angle of ellipse in parentheses;\\
		References: 1 -- blazar monitoring Marscher et al. (VLBA project
codes BM256, S1136); 2 -- dedicated monitoring (VLBA project codes BS193, BS194).\\
		$\dag$ -- not used for the flux density analysis due to gain
calibration problems and bad weather conditions on some of the VLBA antennas.
		\end{flushleft}
	\end{table}

	\begin{figure}
		\resizebox{\hsize}{!}{\includegraphics{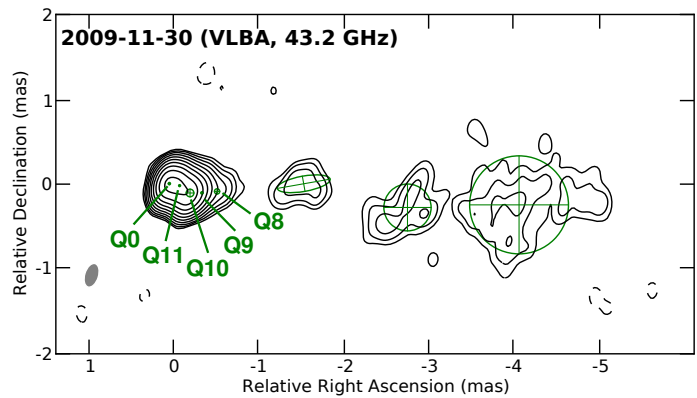}}
		\caption{VLBA image of the total brightness distribution of 3C\,345 at 43.2\,GHz
made from observations on Nov.~30, 2009. Open crossed circles show the FWHM of eight
Gaussian components applied to fit the structure observed. The shaded ellipse, in the
lower left corner, represents the FWHM of the restoring beam. The image peak flux density is
2.1\,Jy\,beam$^{-1}$ and the RMS noise is 1\,mJy beam$^{-1}$. The contour 
levels are (-0.15, 0.15, 0.3, 0.6, 1.2, 2.4, 4.8, 9.6, 19, 38, 77)\,\%
of the peak flux density. Here the nuclear region is modeled by two circular
Gaussian components (Q0, Q11) identified as the best representation of the
observed brightness distribution.}
		\label{fig:map}
	\end{figure}

	The brightness distribution of the radio emission was modelfitted by
multiple Gaussian components providing positions, flux densities and
sizes of distinct emitting regions in the jet. Fig.~\ref{fig:map} illustrates
the observed radio structure and the Gaussian modelfit representation.
In the following, we interpret the eastern-most Gaussian modelfit component 
obtained from the VLBI map, hereafter labeled as Q0 (see Fig.~\ref{fig:map}),
as the base (or ``core'') of the radio jet at 43.2\,GHz. The remaining
features can signify perturbations or shocks developing in the jet. Locations
and proper motions of these jet features are then determined with respect to Q0.

To find the best description of the observed brightness distribution, 
compact emission in the nuclear region ($\leq$\,0.15\,mas from Q0) was modeled
using four different approaches: 1) single circular Gaussian, 2) single
elliptical Gaussian, 3) two circular Gaussians, 4) three circular Gaussians. 
In the following discussion, these modelfitting approaches are designated 1C, 1E, 2C and 3C,
respectively.

The criterion for significant improvement between two modelfits was developed
using a ratio of reduced $\chi^2$ parameters
	\begin{equation}
	  \frac{\chi^2_1}{\chi^2_2} \geq \frac{x_{0.68;2}}{x_{0.68;1}},
	\end{equation}
where the subscripts refer to the two modelfits compared. The order
is defined by the number of degrees of freedom $n$ of the
individual modelfits with the condition $n_1<n_2$. The parameter $x_{0.68}$ is
obtained by solving the equation of the cumulative distribution
function (CDF) of the $\chi^2$ distribution,
	\begin{equation}
	  0.68 = \frac{1}{\Gamma\left(\frac{n}{2}\right)} \,
\gamma\left(\frac{n}{2},\frac{x_{0.68}}{2}\right),
	\end{equation}
where $\Gamma$ denotes Euler's Gamma function and
$\gamma$ represents the lower incomplete Gamma function. The number of degrees
of freedom, $n$, depends on the number of modelfit parameters,
since $\chi^2_\alpha = \chi^2_\mathrm{min} +
x_\alpha\left(n,\alpha\right)$, with $\alpha$ describing the significance
level ($\alpha = 0.68$, in our case). The proof and application of this property
are discussed in \citet{1976ApJ...210..642A}.

\begin{figure}
		\resizebox{\hsize}{!}{\includegraphics{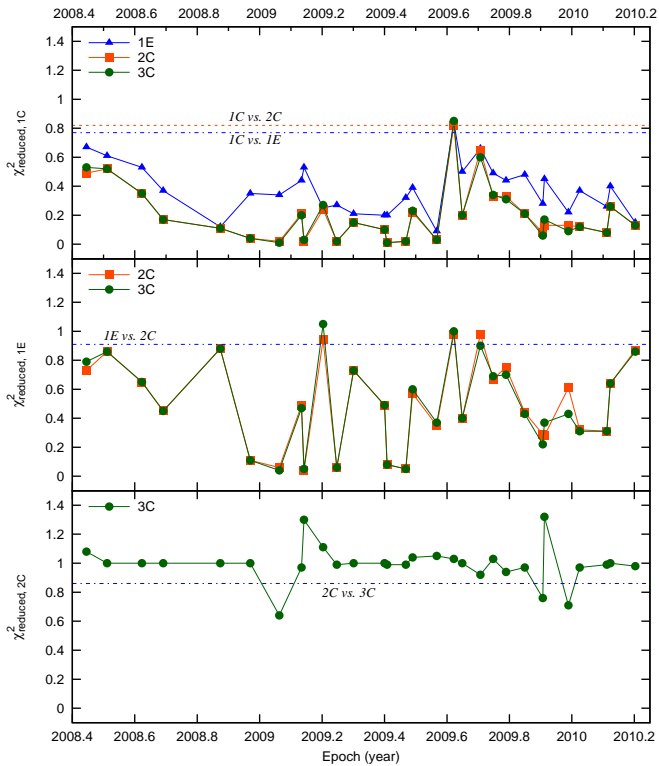}}
		\caption{Reduced $\chi^2$ parameters of the modelfit
representations of the core region. The top panel shows the resulting $\chi^2_\mathrm{reduced}$ values
of 1E, 2C and 3C Gaussian models normalized
to the value of the 1C Gaussian representation. The two horizontal lines mark the 
limit below which, for the corresponding model, a significant improvement (68\% confidence)
over 1C is given (0.77, 0.82). Similarly the middle panel shows the $\chi^2_\mathrm{reduced}$
values for 2C and 3C normalized to that of 1E. The horizontal line 
marks the significant improvement level compared to 2C (0.91). The bottom panel normalizes
the $\chi^2$ values to that of 2C. The dashed line plots the significant improvement 
limit for 3C (0.86).}
		\label{fig:fitscompare}
	\end{figure}

Fig.~\ref{fig:fitscompare} compares the $\chi^2$ parameters of each modelfit,
normalized to the $\chi^2$ values of the 1C (top panel), 1E
(middle panel) and 2C (bottom panel). The horizontal
lines in the plots represent the upper thresholds for
significant improvement over the labeled Gaussian model. The 1C models
compared to the 1E and 2C models have the respective threshold values of 0.77 and 0.82. 
1E models compared to 2C models have a threshold value of 0.91, and
2C compared to 3C models needs to have an improvement of 0.86 or less.
For some of the observations, the representation of the nuclear region by 1E 
is comparable to 2C. However, in order to form a consistent dataset, we adopt the
representation by 2C providing the optimal ratio
of $\chi^2$ to the number of model parameters for all epochs.

	The errors of the obtained modelfit parameters were estimated from the image plane, following 
analytical approximations introduced by \citet{1999ASPC..180..301F} and \citet{2005astro.ph..3225L} with modifications to account
for the strong side-lobe case inherent to VLBI observations. For each component in the model, the root-mean-square (RMS) noise 
($\sigma_{\mathrm{p}}$) was determined
measuring around its position in the residual image. Then the respective component was removed from the Gaussian model and 
the modified model was subtracted from the data, yielding an image that contained only the contribution from the
component investigated. The flux density ($S_\mathrm{p}$) at the peak position of the Gaussian was taken. 
Further input values were, observation beam size ($b = \sqrt{b_\mathrm{maj}\cdot b_\mathrm{min}}$), component total flux density ($S_\mathrm{t}$), 
component distance ($r$), position angle ($\Phi$), and size ($d$). The errors for the
model fit parameters were calculated in the following way:
  \begin{equation}
    SNR = \frac{S_\mathrm{p}}{\sigma_\mathrm{p}}\quad\left(SNR > 1\right)
  \end{equation}
  \begin{equation}
    \sigma_\mathrm{t} = \left(\sigma_\mathrm{p} \cdot\sqrt{1+SNR}\right)\sqrt{1+\left(\frac{S_\mathrm{t}^2}{S_\mathrm{p}^2}\right)}
  \end{equation}
  \begin{eqnarray}
    d_\mathrm{lim} &=& \frac{4}{\pi}\cdot\sqrt{\pi \ln{(2.0)}\cdot b\cdot \ln{\left(\frac{SNR}{SNR-1}\right)}};  \\
                   &&\hspace{15ex} \mathrm{if}\quad (d_\mathrm{lim} > d)\quad\mathrm{else}\quad (d_\mathrm{lim} = d)
  \end{eqnarray}
  \begin{equation}
    \sigma_\mathrm{d} = \frac{\sigma_\mathrm{p}\cdot d_\mathrm{lim}}{S_\mathrm{p}}
  \end{equation}
  \begin{equation}
    \sigma_\mathrm{r} = 0.5\cdot \sigma_\mathrm{d}
  \end{equation}
  \begin{equation}
    \mathrm{if}\,(r>0): \sigma_\Phi = \arctan{\left(\frac{\sigma_\mathrm{r}}{r}\cdot\frac{180}{\pi}\right)}.
  \end{equation}
These error estimates only reflect the statistical image errors and are assumed uncorrelated. No additional systematic 
errors were taken into account, such as errors of the amplitude calibration that could add an additional uncertainty to 
the determined flux density values.  

\section{Results}

\subsection{Evolution of the radio emission in the nuclear region}

	Comparisons of the component positions, flux densities and sizes determined
from the Gaussian modelfits revealed a new moving emission region, labeled Q9, first 
detected in the VLBA image from June 16, 2008, followed by detections of another new 
component on January 24, 2009 (Q10), a third one on July 27, 2009 (Q11), and a fourth on
November 28, 2009 (Q12). In the following, components Q9, Q10, Q11 and Q12 (Fig.~\ref{fig:map})
observed within a distance of 0.5\,mas from the core Q0 are referred to as the ``jet''.

\subsubsection{Kinematics\label{sec:kinematic}}

\begin{figure}
		\resizebox{\hsize}{!}{\includegraphics{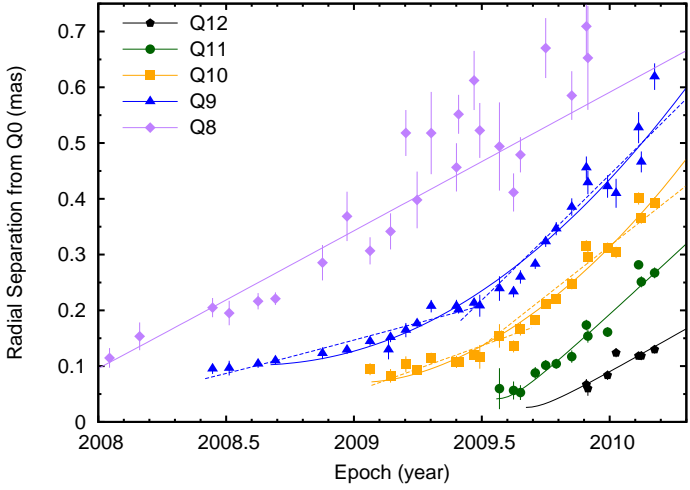}}
		\caption{Evolution of the radial separations from the core of components Q12, Q11,
Q10, Q9 and Q8. Q12-Q9 are related to the radio variability since
2008, Q8 is related to a previous radio flare observed in 2007. The lines are the 
results of polynomial fits to $x(t)$ and $y(t)$ directions separately. The dashed lines
are the results of linear fits to $x(t)$ and $y(t)$ to determine ejection epochs and 
speeds for the two cases discussed in Section~\ref{sec:ejection}.}
		\label{fig:separation}
	\end{figure}

	We determine the kinematics of individual features from their relative positional offsets 
with respect to Q0. The temporal evolution of the measured offsets is plotted in Fig.~\ref{fig:separation}. The component 
Q8, which was first observed in 2007, is also included for the purpose of comparison.

	The motions in the jet of 3C\,345 are investigated using the R.A., Dec. ($x$, $y$) positions 
of a jet component relative to the core component (Q0) over the observed periods, 
fitting them separately using polynomials of different order (cf. \citealt{1995ApJ...443...35Z})
and applying the following procedure:
\begin{enumerate}
 \item The first order polynomial is fitted to the trajectory using 
the nonlinear least-squares Marquardt-Levenberg algorithm (as implemented by \textit{gnuplot}).
 \item The second order polynomial is fitted to the trajectory and if the improvement of the reduced $\chi^2$
satisfies the criterion described in Section~\ref{sec:vlbiobs}, the fit is accepted.
 \item The order of the fit polynomial is increased until no improvement according to the $\chi^2$ statistics 
is achieved.
\end{enumerate}

	Using this approach, we find that it is sufficient to represent the 
trajectories of components Q8, Q11 and Q12 by linear fits in both the $x$ and $y$ directions. 
In the case of components Q9 and Q10, 
a second order polynomial represents the best fit
to the observed data in the $x$ direction, implying apparent acceleration. 
In the $y$ direction a linear fit is sufficient. The resulting fitted radial separations, $r(t) = \sqrt{x(t)^2 + y(t)^2}$,
are drawn in Fig.~\ref{fig:separation}. For each component, the fits yield average proper motion and mean angular
speed $\langle\mu\rangle$ and the average direction of motion $\langle\Phi\rangle$. The kinematic properties thus derived 
for the jet components indicate that Q9 and Q10 underwent a clear phase of apparent acceleration over a period of 1.5 years 
and over a distance of $\sim$0.3\,mas (2\,pc). No statistically significant acceleration was observed for Q8, Q11 and Q12. 
The observed values for $\langle\mu\rangle$ are in the range of 0.25 - 0.42 mas\,year$^{-1}$ and for $\langle\Phi\rangle$ 
in the range of $-96$ to $-120$\degr, also see Table~\ref{tab:kinsummary}. 
\citet{2005AJ....130.1418J} previously reported apparent jet component speeds of 
0.29 -- 0.69 mas\,year$^{-1}$ and a jet position angle of -66 -- -95\degr.  
Using $\langle\mu\rangle$, the average apparent speed $\overline{\beta}_\mathrm{app}$ and the speed in the source 
frame (deprojected) $\overline{\beta}$ are derived through,
  \begin{equation}
    \overline{\beta}_\mathrm{app} = \langle\mu\rangle\,\frac{D_\mathrm{L}}{\left(1+z\right)}
  \end{equation}
and
  \begin{equation}
    \overline{\beta} = \frac{\overline{\beta}_\mathrm{app}}{\overline{\beta}_\mathrm{app}\cos{\Theta} + \sin{\Theta}}\quad,
  \end{equation}
where $D_\mathrm{L}$ is the luminosity distance, $z$ is the redshift and $\Theta$ is the 
jet angle to the line of sight (viewing angle).

As a final step the physical parameters, Doppler factor $\delta$, Lorentz factor $\Gamma$ and
viewing angle $\Theta$ of each jet component are derived. For the following calculations, we assume 
that the jet emission is dominated by radiative losses \citep[see e.g.,][]{2005AJ....130.1418J} and consist 
of optically thin shocked gas, $\alpha=-0.7$ (i.e. $ S_\nu \propto  R^{-1.633}\cdot\nu^{+\alpha}$, 
where $R$ is the size of the emitting region in the rest-frame of the jet and $\nu$ the
frequency in the observer's frame, \citealt{1985ApJ...298..114M}). The variability Doppler factor is then 
derived as,
  \begin{equation}
    \delta_\mathrm{var} = \frac{d_\mathrm{eff} D_\mathrm{L}}{c \Delta t_\mathrm{var} \left(1+z\right)}, 
  \end{equation}
where $D_\mathrm{L}$ is the luminosity distance and $d_\mathrm{eff}$ is the effective angular size
of a spherical region (i.e. the measured FWHM of the component multiplied by 
a factor of 1.8; cf. \citealt{1999ASPC..180..335P}). The flux variability time scale is defined as $\Delta t_\mathrm{var} = dt/\ln(S_\mathrm{1}/S_\mathrm{2})$. 
The value of $S_\mathrm{1}$ is the measured maximum component flux density, while $S_\mathrm{2}$ represents the minimum component
flux density selected at the time of maximum absolute value of the time derivative of the flux density. It is then possible 
to calculate bulk Lorentz factors by combining the 
derived $\delta_\mathrm{var}$ with $\beta_\mathrm{app}$,
  \begin{equation}
    \Gamma_\mathrm{var} = \frac{\beta_\mathrm{app}^2 + \delta_\mathrm{var}^2 + 1}{2\delta_\mathrm{var}}.
  \end{equation}

Combining the derived values for all components, a mean Lorentz factor of 12.5 and a mean Doppler 
factor of 14.4 are obtained (see Table~\ref{tab:kinsummary}).

An upper limit for the viewing angle $\Theta$ can immediately be determined using
the average apparent component speed $\overline{\beta}_\mathrm{app}$ and the relation
$\delta_\mathrm{min} = \sqrt{1+\beta_\mathrm{app}^2}$. This combined with the equation
\begin{equation}
    \label{eqn:theta}
    \Theta = \arctan{\frac{2\beta_\mathrm{app}}{\beta_\mathrm{app}^2 + \delta^2 - 1}}\quad,
\end{equation}
yields $\Theta \leq 5.2\degr$. Using the variability time scale argument and the
derived values of $\delta_\mathrm{var}$ from above, the values $\Theta_\mathrm{var}$ for each 
component are obtained, which together have a mean $\Theta_\mathrm{var}$ of 4.7\degr$^{+0.65}_{-0.51}$, consistent
with our upper limit. 
\citet{2009A&A...507L..33P} obtained a similar value of $\Theta = 5.1\degr$, 
which was determined by combining the component speeds at 15\,GHz and 
variability Doppler factors derived from single-dish observations at 37\,GHz. 
Earlier, \citet{2005AJ....130.1418J} obtained a smaller viewing angle of 
2.7\degr$\pm$\,0.9\degr\ using VLBI data at 43.2\,GHz alone. In the following discussion,
we adopt $\Theta$ = 5\degr.

  \begin{table*}
    \caption{Measured and derived physical parameters for radio emission regions in the inner jet $\leq$0.7\,mas (4.6\,pc).}
    \label{tab:kinsummary}
    \begin{center}
      \begin{tabular}{llllll}
	\hline
	Label	                        & Q8	                       & Q9	                     & Q10	                    & Q11	                   & Q12\\
	\hline
	\#	                        & 26	                       & 30	                     & 23	                    & 13	                   & 7\\
	$<\mu>$ ($\beta_\mathrm{app}$)  & 0.248\,$\pm$\,0.022 (8.5c)   & 0.277\,$\pm$\,0.012 (9.5c)  & 0.300\,$\pm$\,0.011 (10c)    & 0.427\,$\pm$\,0.020 (15c)    & 0.272\,$\pm$\,0.027 (9.0c)\\
	$<\mu_\mathrm{lower}>$          & -                            & 0.1260\,$\pm$\,0.0088       & 0.165\,$\pm$\,0.016          & -                            & -\\
	$<\mu_\mathrm{upper}>$          & -                            & 0.458\,$\pm$\,0.089         & 0.360\,$\pm$\,0.049          & -                            & -\\
	$<\Phi>$                        & -100.0\degr\,$\pm$\,2.9\degr & -96.3\degr\,$\pm$\,1.1\degr & -114.3\degr\,$\pm$\,1.8\degr & -120.5\degr\,$\pm$\,3.5\degr & -117.8\degr\,$\pm$\,5.4\degr \\
        $<\Phi_\mathrm{lower}>$         & -                            & -97.0\degr\,$\pm$\,1.0\degr & -109.7\degr\,$\pm$\,3.2\degr & -                            & -\\
        $<\Phi_\mathrm{upper}>$         & -                            & -96.2\degr\,$\pm$\,1.6\degr & -112.3\degr\,$\pm$\,2.1\degr & -                            & -\\
	$\tau_\mathrm{e}$               & 2007.62\,$\pm$\,0.12         & 2006.91\,$\pm$\,0.11        & 2009.055\,$\pm$\,0.025       & 2009.544\,$\pm$\,0.018       & 2009.651\,$\pm$\,0.042\\
	$\tau_\mathrm{lower}$           & -                            & 2007.834\,$\pm$\,0.083      & 2008.674\,$\pm$\,0.073       & -                            & -\\
	$\tau_\mathrm{upper}$           & -                            & 2009.00\,$\pm$\,0.16        & 2009.11\,$\pm$\,0.11         & -                            & -\\
	S$_1$                           & 0.73\,$\pm$\,0.12            & 3.44\,$\pm$\,0.16           & 2.301\,$\pm$\,0.072          & 1.489\,$\pm$\,0.086          & 2.01\,$\pm$\,0.11\\
	S$_2$                           & 0.257\,$\pm$\,0.078          & 0.311\,$\pm$\,0.038         & 0.816\,$\pm$\,0.044          & 0.786\,$\pm$\,0.057          & -\\
	d$_1$                           & 0.114\,$\pm$\,0.038          & 0.1115\,$\pm$\,0.0088       & 0.1012\,$\pm$\,0.0067        & 0.0620\,$\pm$\,0.0094        & -\\
	d$_2$                           & 0.244\,$\pm$\,0.094          & 0.232\,$\pm$\,0.036         & 0.170\,$\pm$\,0.016          & 0.06\,$\pm$\,0.011           & -\\
	$dt$                            & 1.33                         & 0.715                       & 0.385                        & 0.204                        & - \\
	$\Delta t_\mathrm{var}$         & 1.27\,$\pm$\,0.42            & 0.297\,$\pm$\,0.016         & 0.371\,$\pm$\,0.022          & 0.319\,$\pm$\,0.046          & - \\ 
	$\delta_\mathrm{var}$           & 5.6\,$\pm$\,2.2              & 23.3\,$\pm$\,2.2            & 16.9\,$\pm$\,1.4             & 11.9\,$\pm$\,2.5             & -\\
	$\Gamma_\mathrm{var}$           & 9.8\,$\pm$\,1.4              & 13.61\,$\pm$\,0.93          & 11.6\,$\pm$\,0.5             & 15.1\,$\pm$\,1.0             & -\\
	$\Theta_\mathrm{var}$           & 9.4$^{+2.4}_{-1.9}$          & 1.70$^{+0.34}_{-0.23}$      & 2.97$^{+0.42}_{-0.32}$       & 4.66$^{+0.81}_{-0.69}$          & -\\
	(*)                             & \textit{no}                  &  \textit{yes}               &  \textit{yes}                &  \textit{no}                 &  \textit{no}\\
      \end{tabular}
    \end{center}
    \begin{flushleft}
      {\bf Notes:} 
 	\# -- number of data points;\\
	$<\mu>$ ($\beta_\mathrm{app}$), $<\mu_\mathrm{lower}>$, $<\mu_\mathrm{upper}>$ -- average component speed in mas year$^{-1}$, in parentheses the apparent speed $\beta_\mathrm{app}$, for Q9 and Q10 lower and upper limits of the component speeds;\\
	$<\Phi>$, $<\Phi_\mathrm{lower}>$, $<\Phi_\mathrm{upper}>$ -- average position angle of the component motion in degree, for Q9 and Q10 lower and upper limit;\\
	$\tau_\mathrm{e}$ ($\tau_\mathrm{lower}$, $\tau_\mathrm{upper}$)  -- Ejection epoch/time of zero separation from component labeled Q0, determined through linear extrapolation 
		     using the average component speed and component position at the midpoint, 
		     $t_\mathrm{mid} = 0.5\cdot(t_{\mathrm{max}} - t_{\mathrm{min}})$, of the respective dataset, for Q9 and Q10 lower and upper limits;\\
	S$_1$, S$_2$ -- maximum, minimum (at the time of maximum absolute value of the time derivative) component flux density in units of Jy;\\
	d$_1$, d$_2$ -- circular Gaussian modelfit component FWHM in units of mas at the time of S$_1$, S$_2$;\\
	dt -- time between S$_\mathrm{max}$ and S$_\mathrm{min}$ in units of years;\\
	$\Delta t_\mathrm{var}$ -- time variability factor in units of years;\\
	$\delta_\mathrm{var}$, $\Gamma_\mathrm{var}$, $\Theta_\mathrm{var}$ -- derived quantities in order of the variability Doppler and Lorentz factors, and the viewing angle to the line of sight in degrees\\
	(*) -- indicator whether component was accelerating (\textit{yes}) or not (\textit{no})\\[6pt]
 	The asymptotic standard errors for the fits were
	determined statistically using an implementation of the nonlinear least-squares
	Marquardt-Levenberg algorithm with the position errors as weights. Errors 
	of the derived parameters are 68\% confidence limits determined through
	Monte-Carlo simulations.
      \end{flushleft}
  \end{table*}
	

	\subsubsection{Ejection Epochs\label{sec:ejection}}

With the kinematics determined in Section~\ref{sec:kinematic}, it is possible to estimate the time at which a 
moving jet feature passes the VLBI core at 43\,GHz. These passages are referred to 
as ejection epochs. The ejection epoch also marks the time at 
which a new jet feature begins to contribute to the observed radio emission.

As a first approach to determine the ejection epochs, we assume that the jet is optically 
thin all the way to the core
and the features already travel at the observed average speed while passing through the 
VLBI core. Under these assumptions the ejection epoch is determined by back tracing the 
fitted linear trajectories. The resulting ejection epochs for all features
 are listed in Table~\ref{tab:kinsummary}, as well. This approach provides 
reasonable estimates of the ejection epochs for non-accelerating features, 
while for the apparently accelerating components Q9 and Q10 a different approach 
is required.

Looking at the radial separations of Q9 and Q10 in Fig.~\ref{fig:separation}, the acceleration
is most evident for the time before 2009.6.
An upper limit on the ejection epoch can be determined using a linear fit
to the data points after 2009.5 for Q9 and 2009.4 for Q10 ($\tau_\mathrm{upper}$ in Table~\ref{tab:kinsummary}). 
Similarly, lower limits, $\tau_\mathrm{lower}$, on the ejection epochs of Q9 and Q10 are obtained by
considering only data points before 2009.5 and 2009.4 respectively.

	\subsubsection{Jet Intensity Gradient\label{sec:flux}}

\begin{figure}
		\resizebox{\hsize}{!}{\includegraphics{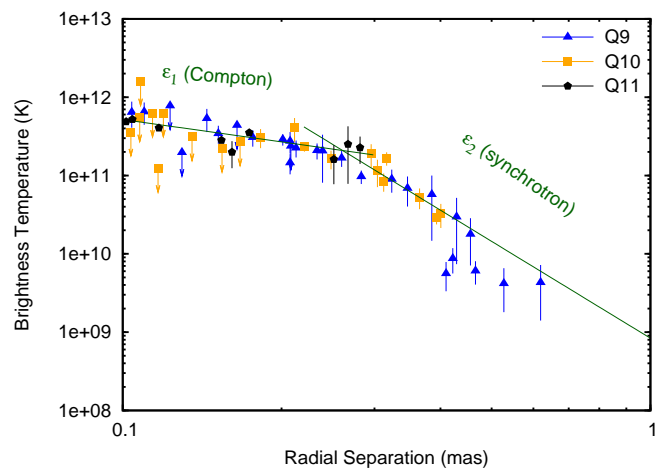}}
		\caption{Component brightness temperatures against radial separation from the VLBI core (Q0), representing
the emission intensity gradient along the jet. Points with arrows are 1$\sigma$ upper limits. 
Two lines are fitted to the data to determine the power law indices $\epsilon$ 
($T_\mathrm{b} \propto d_\mathrm{jet}^{-\epsilon}$) from the data,
the fit from 0.1 to 0.3\,mas yields $\epsilon_1 = 0.95 \pm 0.69$ and 0.3 -- 0.65\,mas yields
$\epsilon_2 = 4.11 \pm 0.85$.}
		\label{fig:Tb}
	\end{figure}

	In flat spectrum radio quasars (FSRQs), such as 3C\,345, the component flux decay is 
	commonly driven by radiative losses \citep{1999ApJ...521..509L, 2005AJ....130.1418J}, 
        which was assumed in Section~\ref{sec:kinematic} without further justification. To test 
        this, the maximum component brightness temperature needs to be calculated as a measure 
for the emission intensity of each component, using:
\begin{equation}
  T_\mathrm{b} = 1.22\cdot10^{12}\cdot\frac{S_\mathrm{comp}\cdot\left(1+z\right)}{d_\mathrm{comp}^2\cdot\nu^2},
\end{equation}
where $S_\mathrm{comp}$ is the component flux density in Jansky, $z$ the redshift of 
the source, $d_\mathrm{comp}$ the FWHM size of the circular Gaussian in mas and 
$\nu$ the observing frequency in GHz. In Fig.~\ref{fig:Tb} the
brightness temperatures for components Q9, Q10 and Q11 are plotted as a function of 
the radial separation from the core.

In the common picture of the shock-in-jet model, a relativistic shock propagates 
down a conical jet, slowly expanding adiabatically albeit maintaining shock conditions.
In this scenario the assumption of a power-law electron energy distribution ($N(E)\,dE \propto E^{-s}\,dE$), a power-law magnetic
field evolution ($B \propto r_\mathrm{jet}^{-a}$) and a constant jet opening angle with the jet transverse
size proportional to the distance along the jet ($d_\mathrm{jet} \propto r_\mathrm{jet} \sin{\Theta}$)
can be made. While the shock continues to travel down the jet, it undergoes three 
major evolutionary stages dominated by Compton, synchrotron and adiabatic energy losses \citep{1992vob..conf...85M}.
From this it follows that the brightness temperature decays as
a power-law, $T_\mathrm{b,jet} \propto r_\mathrm{jet}^{-\epsilon}$, where
$r_\mathrm{jet}$ is the distance in the jet at which $T_\mathrm{b,jet}$ is measured. 
The value of $\epsilon$ can be derived from spectral evolution of radio emission \citep{1999ApJ...521..509L},
assuming $T_\mathrm{b} \propto S_\mathrm{comp}/(r_\mathrm{jet}^2\nu^2)$ and the Doppler factor $\delta = const$.

For Compton ($\epsilon_\mathrm{c}$), synchrotron ($\epsilon_\mathrm{s}$) and adiabatic ($\epsilon_\mathrm{a}$) 
losses, $\epsilon$ are calculated as follows,  
\begin{eqnarray}
  \epsilon_\mathrm{c} &=& \left[\left(s+5\right)+a\left(s+1\right)\right]/8,\\
  \epsilon_\mathrm{s} &=& \left[4\left(s+2\right)+3a\left(s+1\right)\right]/6\\
  \epsilon_\mathrm{a} &=& \left[ 2 \left(2s+1\right) + 3a \left(s + 1\right) \right]/6
\end{eqnarray}
with a typical value of $s = 2.0$ (corresponding to a synchrotron spectral index $\alpha$ = -0.5; $ S_\nu\propto\nu^{+\alpha}$) 
and $a = 1$ (dominant transverse magnetic field), $\epsilon_\mathrm{c} = 10/8 = 1.25$, $\epsilon_\mathrm{s} = 25/6 \approx 4.17$
and $\epsilon_\mathrm{a} = \boldsymbol{19}/6 \approx 3.17$. For the slopes to be shallower than these derived values, 
acceleration resulting in increasing Doppler factors of $\delta \propto r_\mathrm{jet}^b$ may be considered. 
For a moderate acceleration (i.e. $b=0.1-0.2$), a value $\epsilon_\mathrm{c} \approx 0.13$ is obtained
assuming $s=2$ and with a longitudinal magnetic field ($a=2$).

The brightness temperature gradient along the jet, shown in Fig.~\ref{fig:Tb}, reveals 
a possible broken power-law behavior of the jet intensity gradient with a break 
distance of $\sim$0.3\,mas and the two slopes $\epsilon_1 = 0.95\pm0.69$
and $\epsilon_2 = 4.11 \pm 0.85$. The value of $\epsilon_1$ is consistent with 
$\epsilon_\mathrm{c}$, with the possible indication of a mild change in $\delta$, and $\epsilon_2$ 
is consistent with the derived $\epsilon_\mathrm{s}$. However, at the $\sim 1\sigma$ level, 
$\epsilon_2$ is also consistent with $\epsilon_\mathrm{a}$. Using the spectral evolution of a jet 
component, \citet{1999ApJ...521..509L} found evidence for a change from the synchrotron 
to the adiabatic stage at a distance of 1.2--1.5\,mas from the core. This suggests
that the transition from Compton to synchrotron stages is indeed observed. Future 
investigations on the spectral evolution of the jet are able to provide additional
constraints on the transition regions between the Compton, synchrotron, and adiabatic stages.

	\subsection{Radio-$\gamma$-ray correlation\label{sec:longterm}}

	\begin{figure}
		\resizebox{\hsize}{!}{\includegraphics{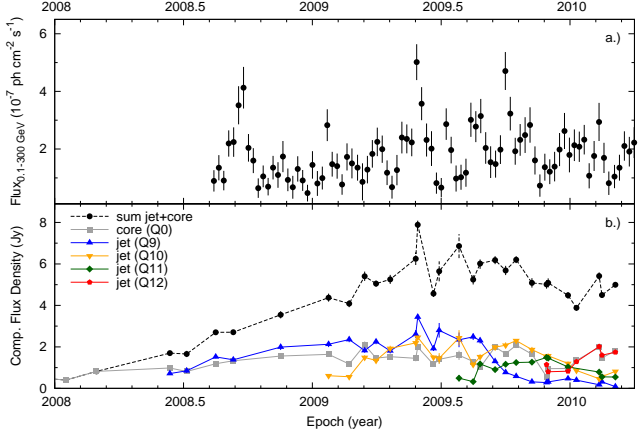}}
		\caption{ \textit{top}: \textit{Fermi} LAT 7-day binned
$\gamma$-ray light curve of 3C\,345 for the energy range of 0.1-300\,GeV.
\textit{bottom:}
VLBA 7~mm component flux densities for the modelfitted VLBI core and inner jet,
represented by up to five circular Gaussian (Q0, Q12, Q11, Q10, Q9). The
component labeled Q0 is the east-most component (see Fig.~\ref{fig:map}) and
represents the compact ``core'' or base of the jet. The black curve plots
the sum of the flux densities of all four components.}
		\label{fig:lightcurve}
	\end{figure}

	The correlation between radio and optical emission was studied in \citet{Schinzel_2010A} for
identification of $\gamma$-ray emission from 3C\,345. Here the connection between the synchrotron
radio emission and $\gamma$-ray emission is investigated. To quantify a possible connection, the
correlation between the $\gamma$-ray and the radio light curve is calculated. For this we used the 
discrete correlation function presented by \citet{1988ApJ...333..646E}, which works well for unevenly sampled data. 
No statistically significant correlation is found, although around 2009.4,  
a rapid change in VLBI flux densities by $\sim$1.5\,Jy within only 4 days coincided
well with a $\gamma$-ray flare (also see Fig.~\ref{fig:lightcurve}).

The top panel of Fig.~\ref{fig:lightcurve} plots the weekly averaged $\gamma$-ray 
light curve of 3C\,345. From it a seemingly long-term trend is seen, most 
evident between 2008.7 and 2009.4. After 2009.4 an increase in short-time variability
is observed, followed by a slightly decreasing trend after 2009.8. The measured $\gamma$-ray flux increases 
by a factor of 2-3 between 2008.6 and 2009.5. Even more evident 
is a long-term trend in the light curve of the radio jet features (bottom panel of Fig.~\ref{fig:lightcurve}), 
following from the sum of the flux density of individual jet components. The optical R band data presented in 
\citet{Schinzel_2010A} similarly showed a rising trend for the same time period.
This dataset is added for comparison as well.

To quantify this relation, the light curves are first rescaled with the zero flux
representing their respective mean value. The radio flux density of the jet has 
a mean value of 3.60\,Jy and corresponds to the emission from the jet of an apparent 
size of $\leq 2$~pc (excluding the core). The core has a mean flux density of 1.5~Jy
and an apparent size of $\sim 0.3$~pc. Note, the jet flux density is by a factor of 2.4 stronger 
than that of the core. The average $\gamma$-ray flux is 
$1.8\cdot 10^{-7}$~ph~cm$^{-2}$~s$^{-1}$. Observations at optical R band
have an average magnitude of 16.6, which corresponds to 
a flux density of 7.7$\cdot10^{-4}$\,Jy.

\begin{figure}
		\resizebox{\hsize}{!}{\includegraphics{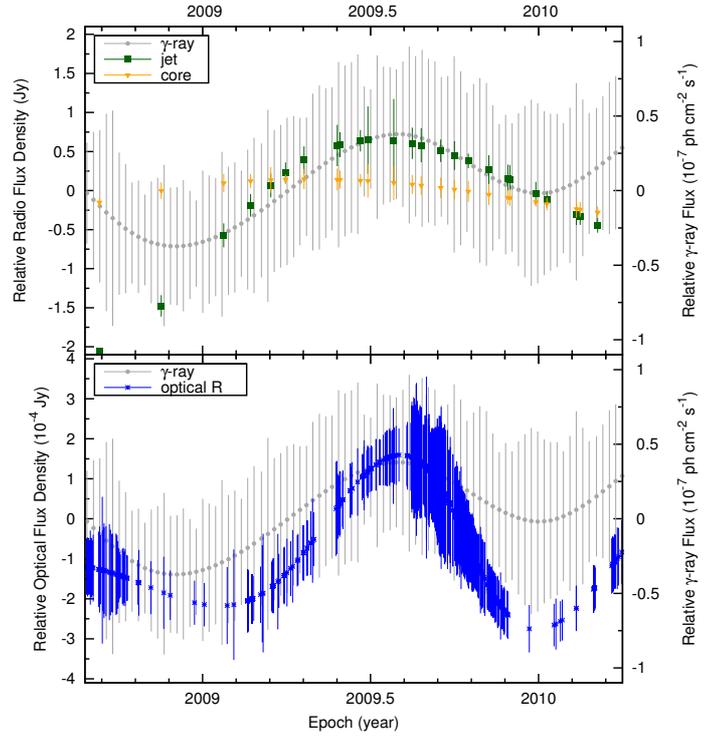}}
		\caption{Long-term trends of the radio jet, 
radio core flux densities, optical R band flux (Schinzel et al. 2010) and the $\gamma$-ray flux relative to their respective
mean values (radio jet: 3.6\,Jy, radio core: 1.5\,Jy, optical R: 16.6$^{m}$ ($7.7\cdot10^{-4}$\,Jy), $\gamma$-ray: $1.8\cdot
10^{-7}$~ph~cm$^{-2}$~s$^{-1}$). The trend was extracted fitting cubic splines
with 0.4 year bins. The relative radio flux density was scaled by a factor of 2:1 with respect to the relative $\gamma$-ray flux, 
the optical flux density was scaled by a factor of 4:1, and the data points match the sampling
of the original light curves.}
		\label{fig:detrend}
	\end{figure}

In the final step, the rescaled light curves are de-trended using cubic spline 
interpolations over 0.4~year bins \citep{1992nrfa.book.....P}. The extracted 
long-term trends are shown in Fig~\ref{fig:detrend}. From these it is 
immediately evident that a similar trend is obtained for the radio jet
and the $\gamma$-rays, whereas the core does not show a significant trend.
The discrepancy of the first 0.4~year is caused by the lack of $\gamma$-ray monitoring 
before 2008.6 and a flare immediately after 2008.6 (see Fig.~\ref{fig:lightcurve}).
The trend of the optical emission shows a similar 
behavior compared to the $\gamma$-ray and radio trends, however it is 
more peaked. Comparing trend amplitudes, the amplitude of the variation in the optical is 4 times 
greater than in the $\gamma$-rays. The radio shows a factor of 2 higher amplitude 
of the variation than $\gamma$-rays. Based on the peak values of these 
trends, the radio leads the $\gamma$-ray trend by 31$^{+29}_{-11}$\,days and 
optical leads the $\gamma$-ray trend by 1.1$^{+11.3}_{-7.7}$\,days. Altogether
this is consistent with an almost zero time lag between the observed long-term trends.

The observed emission from the radio jet showing the matching trend is most of 
the time dominated by the jet component closest to the core. At the beginning
of the monitoring period this is Q9 \& Q10 and by the end of 2010 they are replaced
by Q11 \& Q12. This connects the emission from the radio jet with that observed 
at optical and $\gamma$-ray energies and places the site of the underlying 
multi-wavelength emission within the resolved 43\,GHz radio jet. The very region 
that was shown to be Compton loss dominated in Section~\ref{sec:flux}.

	\section{Discussion} 

	Monitoring of the fine structure of the jet in 3C\,345 reveals a wealth
of morphological and kinematic features of the flow. In particular, compelling evidence 
for a stationary feature in the parsec-scale radio jet of 3C\,345 is found. 
This stationary feature is located at a distance of $\sim$0.1\,mas ($\sim$\,0.7\,pc) from the core.
The kinematics of a large sample of relativistic jets, obtained through the MOJAVE survey\footnote{https://www.physics.purdue.edu/astro/mojave/},
revealed stationary features as a frequent occurence \citep{2009AJ....138.1874L}. 
In particular, in the jets of $\gamma$-ray blazars, observed at 22 and 43\,GHz, stationary features
within 2\,pc (projected) from the core are commonly observed \citep{2001ApJS..134..181J}. 
Stationary jet features were investigated in more detail in the cases of
3C\,390.3 \citep{2010MNRAS.401.1231A}, a nearby double-peaked radio galaxy at $z = 0.0561$, and the 
Seyfert galaxy 3C\,120 ($z = 0.033$; \citealt{2010ApJ...715..355L}). For these two sources,
stationary features were found at distances of $\sim$\,0.4\,pc (3C\,390.3) and 
$\sim$\,1.3\,pc (3C\,120) from the core at 15\,GHz. Their physical origin was 
associated with that of standing shocks. A more detailed investigation 
of the spectral evolution of a radio flare in CTA\,102, a highly polarized quasar at $z$=1.037, 
provides supporting evidence for a shock-shock interaction scenario, providing a 
physical description for stationary features interacting with traveling shocks 
\citep{2011A&A...531A..95F}.

The moving plasma condensations (components) are most likely generated near the 
base of the jet of 3C\,345, but can only be reliably monitored after they 
pass through the location of the stationary feature. The nucleus of the jet 
lies at a distance of $\sim$7\,pc upstream from the 43\,GHz VLBI core \citep{1998A&A...330...79L, PhD...Schinzel2011}. After the passage
of the stationary feature, the superluminal components in 3C\,345 undergoe an apparent acceleration from 
$\sim$\,5 to $\sim$\,15~c over a distance of 0.3\,mas 
in extent ($\sim$23\,pc deprojected). This acceleration seems to be 
a commonly observed feature of many objects in the MOJAVE sample, where a statistically significant 
tendency for acceleration in the base of jets is found \citep{2009ApJ...706.1253H}.  
Previously, \citet{2005A&A...431..831L} discussed the observed acceleration
of a different component in 3C\,345, testing the case of a substantially curved three-dimensional path, causing
the observed dramatic change of the apparent component speed. They concluded that the 
observed acceleration is not likely to be a geometrical effect, but should 
reflect physical acceleration of plasma. The possibility of intrinsic acceleration 
of jet features on parsec scales was discussed by \citet{2004ApJ...605..656V}.
There it was specifically argued that the acceleration observed in 3C\,345 is not purely hydrodynamic, but 
can be attributed to magnetic driving. They showed that the observed acceleration is 
consistent with an acceleration from $\Gamma\sim5$ to $\Gamma\sim10$ over a linear distance 
of $\sim$3 to $\sim$20\,pc. The results presented here are consistent with these 
findings.

The times of passage of knots Q10, Q11, and Q12 through the 43\,GHz VLBI
core coincide with short-term increases in the $\gamma$-ray flux (see
Fig.~\ref{fig:lightcurve}, with times listed in Table~\ref{tab:kinsummary}).
In a similar case, PKS\,1510-089 showed, in addition to the appearance of a new VLBI component, 
a rapid rotation in the optical linear polarization just prior the components passage 
through the core. This was explained by the feature following a spiral path through 
a toroidal field until it crosses a standing shock in the 43\,GHz core \citep{2010ApJ...710L.126M}.

A $\gamma$-ray flare centered on 2009.4 may be related to the brightening 
of radio structure at distances of 0.12 -- 0.20\,mas from the core (0.8 -- 1.3\,pc), 
corresponding to a deprojected distance of $\sim$\,9-15\,pc. Interestingly, this includes
the region for which the presence of a standing shock is suggested by the observations presented
in Sec.~\ref{sec:kinematic}. Radio flares in the compact jet downstream of the VLBI
core are known to occur \citep[e.g., ][]{2001ApJS..133..297W}. Furthermore, \citet{2001ApJS..134..181J,
Jorstad_FmJ_2010} have found that $\gamma$-ray flares often occur after superluminal
knots have separated from the VLBI core. The results presented here support the inference that $\gamma$-ray
emission still occurs parsecs downstream of the central engine, contrary to conclusions
based on theoretical arguments (e.g., \citealt{2010MNRAS.405L..94T, 2010ApJ...717L.118P}).
These arguments place the $\gamma$-ray emission zones within the broad-line region (BLR)
at distances of $\lesssim$1\,pc from the central engine. Within this region a high number of optical--UV 
photons are produced \citep[e.g.][]{2007ApJ...659..997K}, providing ideal
conditions for inverse Compton scattering with relativistic electrons from the jet 
\citep[e.g.][]{2011arXiv1104.4946A, 1997ApJS..109..103D}. In addition, the observed variability
time scales place constraints on the size of the emitting region due to causality arguments.
Usually this size is related to the jet cross section, implying that the emission region is located
close to the black hole. However, these standard assumptions do not necessarily need
to be true.

In Section\,\ref{sec:longterm} we have shown that long-term trends observed
at $\gamma$-ray and optical energies correlate with those of the radio emission at VLBI scales, which
imply a common emission region. This emission region is related to the inner jet,
extending over a distance of up to 23\,pc from the VLBI core, which is at a distance of $\gg$1\,pc from
the central engine, well beyond the BLR. The mechanism for the emission at 
radio and optical wavelengths is therefore dominated by the synchrotron emission 
from the relativistic jet itself. The simultaneity of $\gamma$-ray emission implies
inverse Compton upscattering of synchrotron photons produced in the jet. The 
relative increase of optical and radio emission compared to $\gamma$-ray emission
is a factor of 2 higher in the radio and even a factor of 4 higher in the optical emission, which
is consistent with the mechanism of synchrotron self-Compton. An external contribution
of ``seed'' photons, as proposed by external Compton scenarios \citep[e.g.][]{2010MNRAS.405L..94T} is not required to produce
the amount of the observed $\gamma$-ray emission. Additionally, we have found 
that the properties of the emission in this region are consistent with the Compton-loss 
stage of a shock (Section~\ref{sec:flux}). Altogether, this contradicts the common 
theoretical high energy production scenario described in the previous paragraph. Recent 
observations of TeV emission from FSRQs \citep[e.g.][]{2011ApJ...730L...8A} should 
not have been possible if the high energy emission site were within the BLR, which is expected
to be opaque to $\gamma$-rays at TeV energies due to $\gamma\gamma$ interactions.

The observed kinematics of the jet is now combined with the rise times of 
the radio, optical, and $\gamma$-ray emission. The component Q9 is the first new 
component identified in 2008 and is most likely the emission region responsible 
for the initial flux increase. Using the long-term trends extracted, the 
distance at which Q9 was located during the onset of the flaring period 
in 2008 is $\sim$0.05\,mas from the VLBI core. This marks the region
between the stationary feature and the core at 43\,GHz. Instead of using the 
combined flux trend that is related to the sum of the radio emission regions,
we calculate the time between the peak brightness of Q9 (2009.40) and the first
$\gamma$-ray flare observed (2008.73). If we use the average component speed and 
assume no acceleration, then the closest Q9 could have been during the time 
of the first observed $\gamma$-ray flare is consistent with the position of the 
radio core at 43~GHz. This finding strongly supports the interpretation of the 
parsec-scale jet being responsible for driving the observed high energy emission.

\citet{2011ApJ...726L..13A} have reported on the location of a $\gamma$-ray flare in OJ\,287, 
which was placed at a distance of more than 14\,pc from the central engine. They proposed a model 
for the multi-wavelength emission where the high energy emission could be explained by the synchrotron 
self-Compton process or inverse Compton scattering of infrared radiation from a hot dusty torus, although 
it was concluded that the hot dusty torus scenario is less likely. In the case of 3C\,345 it seems to be 
clear that multiple compact emission features are responsible for the observed variability and what seems to 
be a reasonable model for the case of OJ\,287, could apply for 3C\,345 as well. A new plasma disturbance passes 
through a first conical shock in the core and produces a fast rise in high energy emission; as it continues to propagate
down the jet it continues to produce high energy emission. 

	\section{Summary \& Conclusions}

	Based on 32 VLBA observations of 3C\,345 at 43.2\,GHz, we have investigated
the structure and evolution of the radio emission and related this to variations
 of the $\gamma$-ray and optical emission observed 
2008--2010 by \textit{Fermi}-LAT and a number of optical observatories.

	We have identified and analyzed four new moving emission features
(jet components) in the radio jet of 3C\,345. These regions are found
to move at apparent speeds of 9--15\,c, with the corresponding Doppler
and Lorentz factors derived to be in the ranges of 12--23 and 12--15,
respectively. The kinematic data strongly favor a viewing angle of
4.7\degr\ between the jet axis and the line of sight.

	We have presented evidence for the $\gamma$-ray emission to be produced not
in a compact region near the central engine of the AGN, but in the
Compton-loss dominated zone of the parsec-scale jet extending up to
$r\approx 0.3$\,mas, corresponding to a deprojected linear extent of
$\approx$\,23\,pc (accounting for the source distance and jet orientation). This
zone is further marked by a break in the evolution of the brightness
temperature, $T_\mathrm{b}$, of the radio emission, with $T_\mathrm{b}
\propto r^{-(0.95\,\pm\,0.69)}$ at $r\le 0.3$\,mas and $T_\mathrm{b} \propto r^{-(4.11\,\pm\,0.85)}$ 
at larger separations. Ejections of new superluminally moving and
apparently accelerating features in the jet are linked to the flaring component
of the $\gamma$-ray and optical emission.

These findings favor the synchrotron self-Compton mechanism of the
high-energy emission production, while questioning the entire class of
models that place the high energy emission site within 1\,pc from the
central engine of AGN, with external seed photons for inverse Compton
scattering from the accretion disk or broad-line-region.  They also imply that a significant part of
the $\gamma$-ray emission is generated by highly-relativistic
electrons propagating at a large bulk speed in the jet. In the context
of newly emerging results, stimulated by Fermi data, more detailed
analytical and numerical descriptions of such a scenario are clearly needed
in order to explain the observed connection between radio and
$\gamma$-ray variability. At the same time, continued monitoring and more densely sampled VLBI
observations of well studied, bright blazar radio jets represent an
essential requirement for improving the degree of detail and
statistical accuracy of the correlation reported and further improving
the spatial localization of individual flares in relativistic jets. 

	\begin{acknowledgements}
	We thank Nicola Marchili for providing us with his
implementation of the cubic spline de-trending algorithm.  We thank the anonymous
referee for valuable comments. Frank Schinzel was supported for this 
research through a stipend from the International Max-Planck 
Research School (IMPRS) for Astronomy and Astrophysics at the Universities of 
Bonn and Cologne. We thank NASA for support under FERMI grant GSFC \#21078/FERMI08-0051. 
The research at Boston University was supported in part by NASA through Fermi Guest 
Investigator grants NNX08AJ64G, NNX08AU02G, NNX08AV61G, and NNX08AV65G, and by
National Science Foundation grant AST-0907893. The National Radio Astronomy Observatory 
is a facility of the National Science Foundation operated under cooperative agreement by
Associated Universities, Inc. This research has made use of the NASA/IPAC Extragalactic
Database (NED) which is operated by the Jet Propulsion Laboratory, California
Institute of Technology, under contract with the National Aeronautics and Space
Administration. This research has made use of NASA's Astrophysics Data System. 
	\end{acknowledgements}

	\bibliographystyle{aa}
	\bibliography{references}
	
\end{document}